\documentclass{article}

\usepackage{dcase2020,amsmath,graphicx,url,times,booktabs, tabularx}
\usepackage{hyperref}
\usepackage{subcaption}
\usepackage{xcolor}
\usepackage{multirow}
\usepackage{makecell}
\usepackage{bbm}
\usepackage{dsfont}
\usepackage{enumitem}
\usepackage{float}

\title{ Anomalous Sound Detection as a Simple Binary Classification Problem
with Careful Selection of Proxy Outlier Examples
}

\name{Paul Primus$^1$, Verena Haunschmid$^1$, Patrick Praher$^3$, and Gerhard Widmer$^{1,2}$}
\address{
$^1$Institute of Computational Perception  \\ 
$^2$LIT Artificial Intelligence Lab \\
Johannes Kepler University, Austria \\
$^3$ Software Competence Center Hagenberg GmbH, Austria}

\begin{document}

\ninept
\maketitle

\begin{sloppy}

\begin{abstract}
Unsupervised anomalous sound detection is concerned with identifying sounds that deviate from what is defined as ``normal'', without explicitly specifying the types of anomalies. A significant obstacle is the diversity and rareness of outliers, which typically prevent us from collecting a representative set of anomalous sounds. As a consequence, most anomaly detection methods use unsupervised rather than supervised machine learning methods. 
Nevertheless, we will show that anomalous sound detection can be effectively framed as a supervised classification problem if the set of anomalous samples is \emph{carefully substituted} with what we call \emph{proxy outliers}. Candidates for proxy outliers are available in abundance as they potentially include all recordings that are neither normal nor abnormal sounds. We experiment with the machine condition monitoring data set of the 2020's DCASE Challenge and find proxy outliers with matching recording conditions and high similarity to the target sounds particularly informative. If no data with similar sounds and matching recording conditions is available, data sets with a larger diversity in these two dimensions are preferable.
Our models based on supervised training with proxy outliers achieved rank three in Task 2 of the DCASE2020 Challenge.

\end{abstract}

\begin{keywords}
Unsupervised Anomaly Detection, Proxy Outliers, Outlier Exposure, Anomalous Sound Detection, Machine Condition Monitoring, DCASE2020
\end{keywords}

\section{Introduction}

Automatic detection of anomalies in audio signals is an active field of research in machine learning. This class of problems has various applications in diverse domains, such as novelty-detection in music~\cite{Lu16}, audio surveillance of public spaces~\cite{lim2017}, or acoustic Machine Condition Monitoring (MCM) for predictive maintenance~\cite{Koizumi_2019_0}. 

In this work, we elaborate on our findings for Task 2 of 2020's IEEE DCASE Challenge \cite{Koizumi_2020,Primus2020}, which is concerned with unsupervised MCM. MCM systems aim to detect sounds that deviate from what is considered ``normal'' for a specific machine or a class of machines. Since anomalies (i.e., outliers) typically are rare and diverse, defining and collecting all possible variants in a sufficient quantity to train a classifier is hardly feasible. Moreover, collecting anomalous sounds in MCM often means damaging or destroying the machine, resulting in undesired costs. One workaround is to learn models without explicitly specifying and collecting anomalies, which is commonly referred to as unsupervised Anomaly Detection (AD). However, in this more general scenario, no guidance through anomalous data is available, which makes generalization to future outliers hard.


Most studies in the field disregard data that is unrelated to the task. However we will show in this work that unrelated data, \emph{if chosen carefully}, can be used as a substitute for unavailable anomalous data (Fig.~\ref{fig:data_kinds}). This consequently allows us to frame the unsupervised AD problem as a regular classification problem. We call samples from the unrelated data that will be used for AD \emph{Proxy Outliers} (PO). Our approach is supported by the recommendation for outlier analysis by Aggarwal~\cite{Aggarwal2017} who suggests to  ``always use supervision where possible''. In the experimental section, we will investigate under which conditions unrelated sounds are informative for anomaly detection and finally compare our results to unsupervised baselines based on reconstruction error and density estimation.

By merely training a ResNets to distinguish between normal machine sounds and carefully selected proxy outliers, we achieved the 3rd team-rank in Task 2 of the 2020's DCASE Challenge (Unsupervised Detection of Anomalous Sounds for Machine Condition Monitoring). This is especially remarkable as we did not apply data augmentation, transfer learning, model ensembling, or sophisticated postprocessing. Our implementation is available on GitHub \footnote{\href{https://github.com/OptimusPrimus/dcase2020_workshop}{{https://github.com/OptimusPrimus/dcase2020\_workshop}}}.

\begin{figure}[t!]
    \centering

        \includegraphics[width=0.2\textwidth]{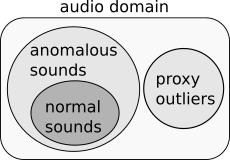}
    \caption{Venn diagram of the data categories. Normal and anomalous sounds are emitted by the target machine in normal and anomalous state, respectively. Proxy outliers are carefully selected from unrelated data in the audio domain.}
    
    \label{fig:data_kinds}
\end{figure}

\section{Related Work}

According to Chandola et al.~\cite{Chandola2009}, anomaly detection (AD) is the task of identifying patterns in the data that differ from what is regarded as normal and, in contrast to noise, are of interest for various downstream tasks. Following Aggarwal's taxonomy~\cite{Aggarwal2017}, we distinguish between the supervised and unsupervised  \emph{AD settings} based on whether the characters of anomalies are well defined or unclear. In \emph{supervised AD}, both normal and anomalous data are defined, available, and labeled for training; the learning task is to fit a classifier that generalizes to future data. Unfortunately, due to the variety of anomalies, it is often hard, in practice, to define and collect anomalies. Even if representative sampling is possible, the class imbalance due to the rare nature of anomalies makes learning particularly challenging. On the other hand, \emph{unsupervised AD} does not make any prior assumptions about the nature of anomalies and is therefore applicable in broader scenarios such as machine condition monitoring . The learning objective turns into creating a model of the ``normal'', which is then used to assign anomaly scores to new samples at inference time (e.g., based on log-likelihoods).

Pimentel et al.~\cite{Marco2014} categorize AD methods into probabilistic,  reconstruction-based, distance-based, domain-based, and information-theoretic approaches. The baseline methods we will apply in our work fall into the first two categories and are both based on neural networks. \emph{Reconstruction-based methods} are typically based on Autoencoders or Variational Autoencoder~\cite{KingmaW13} trained to minimize the reconstruction error on the normal training set (e.g.,~\cite{Koizumi_2019_0, MarchiVESS15}). At test time, samples are encoded and decoded using parameters tailored for normal samples, which usually yields a higher reconstruction error for novel patterns in anomalous data. Note that this method also yields small reconstruction errors for outlier sounds if the Autoencoder generalizes to the anomalous data. Early \emph{probabilistic models} for AD were based on parametric and non-parametric density estimation methods like Gaussian mixture models and kernel density estimation~\cite{Marco2014}. Recent work in this category uses Normalizing Flows~\cite{Schmidt2019_NoveltyDetection} to learn a tractable likelihood approximation of the training data. However, Kirichenko et al.~\cite{Kirichenko2020} show that the latent representations of normalizing flows are mostly based on local pixel correlations, leading to undesired large likelihood values for semantically unrelated samples.

Recent work in unsupervised AD leverages large amounts of unrelated image data, which is commonly referred to as \emph{outlier exposure} \cite{Hendrycks, Ruff_2020}. Most related to our work are the findings of Ruff et al. \cite{Ruff_2020}, who report that a classifier trained on normal data and only 64 random unrelated images can outperform the current state of the art in deep AD. In their experiments, they attribute the unrelated dataset's informativeness to the diversity of the content and the rich multi-scale structure of images.

\section{Method}

\label{sec:OEC}

 To overcome the scarcity problem of anomalous sounds, we propose to substitute real outliers with \emph{Proxy Outliers} (PO), i.e., \emph{carefully selected} recordings that are neither normal nor abnormal sounds. (Fig.~\ref{fig:data_kinds}). Note that, compared to anomalous sounds, POs are cheap and easy to collect if not already available in abundance. In the experimental section, we test a variety of candidate PO sets and investigate under which conditions such arbitrary recordings can be used as POs. To show that our approach can be applied in an unsupervised AD setting (i.e., without explicitly defining anomalies), we make no assumptions on the nature of anomalies and presume only normal data and various candidates for POs are given.

To determine what kind of POs can be utilized for acoustic MCM, we take advantage of machine sounds contained in the combined version of the MIMII \cite{Purohit_2019} and ToyAdmos \cite{Koizumi_2019} data sets released for Task 2 of the DCASE2020 Challenge \cite{Koizumi_2020}. This set includes sounds of 41 machine instances, categorized into six different machine types: fan, pump, slider, toy car, toy conveyor, and valve. We train deep anomaly detectors per machine instance, using the normal sounds of each instance as target data set and several combinations of the remaining machines instances' training sets as POs. Note that the test sets of MIMII and ToyAdmos contain real anomalous machine sounds, and are therefore exclusively used to evaluate our approach. To further increase the variety of PO candidates, we also experiment with the TAU Urban Acoustic Scenes 2020 Mobile (TAU-UAS) data set \cite{Mesaros2018} and the balanced version of AudioSet \cite{AudioSet}, which contains sounds of humans, animals, and even music.  \\

\subsection{Normal Sound Data Sets}

The ToyAdmos and the MIMII data set contain recordings of several machines in a normal and an anomalous operation state, which are grouped into two and four different machine types, respectively. For each machine type, recordings of seven machine instances are available, except for toy conveyor where only six instances are given. Following the DCASE Challenge, we split the data of each machine instance into a training and a testing set, such that the first one contains only normal sounds, and the second one contains both normal and anomalous sounds. To train an anomaly detector for a specific target machine instance, we use the training set of this particular machine instance as normal data. Note that for three instances of each machine type, the test data labels are not given, since these are used to rank the DCASE Challenge submissions. Therefore, we use the training data of these instances only as part of the proxy outlier data. For further details on the exact recording procedure and the distribution of train and test data per machine instance, we would like to refer the reader to the original data set papers \cite{Koizumi_2019, Purohit_2019} and the DCASE Challenge description \cite{Koizumi_2020}.

\subsection{Proxy Outlier Set Candidates}

Having defined the normal data set for each machine instance, we will now describe how the corresponding six candidate PO sets are created from the remaining machines' normal sets, TAU-UAS, and AudioSet.

First, we create three combinations of the remaining machines' normal data so that recording conditions are preserved. Note that ToyAdmos and MIMII were recorded using separate hardware and following a different procedure, and therefore must not be used simultaneously for training if identical recording conditions are desired. For a particular target machine instance in ToyAdmos (MIMII), we consequently combine only the remaining machines' normal data in ToyAdmos (MIMII). The three PO variants (Fig.~\ref{fig:combinations}) are obtained by combining
\begin{enumerate}[label=(\alph*)]
    \item the remaining training sets of the \emph{same} machine type,
    \item the remaining training sets of \emph{different} machine types, or
    \item \emph{all} remaining training sets.
\end{enumerate} 

Furthermore, we experiment with three PO candidate sets with mismatching recordings conditions, specifically with \begin{enumerate}[label=(\alph*)]
    \setcounter{enumi}{3}
    \item the training sets of ToyAdmos if the target instance is in MIMII and vice versa,
    \item the TAU-UAS data set, and
    \item the balanced version of AudioSet.
\end{enumerate}
Note that AudioSet and TAU-UAS are both rich in recording conditions and content, while ToyAdmos and MIMII only contain machine sounds and follow a rigorous recording procedure.

\begin{figure}[t]
    \centering
    \begin{subfigure}{.15\textwidth}
        \centering
        \includegraphics[width=1\textwidth]{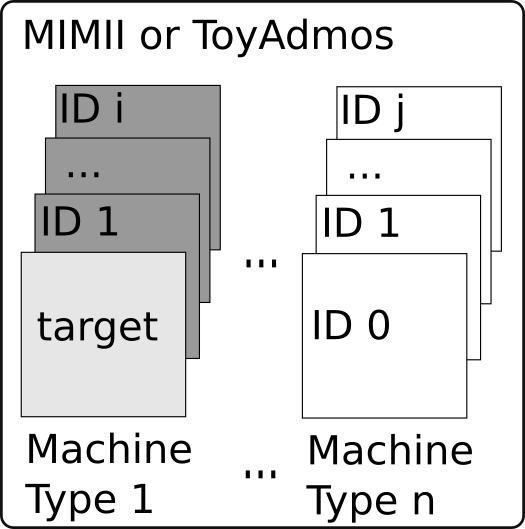}
        \caption{Same Type}
        \label{fig:same}
    \end{subfigure}
    \begin{subfigure}{.15\textwidth}
        \centering
        \includegraphics[width=1\textwidth]{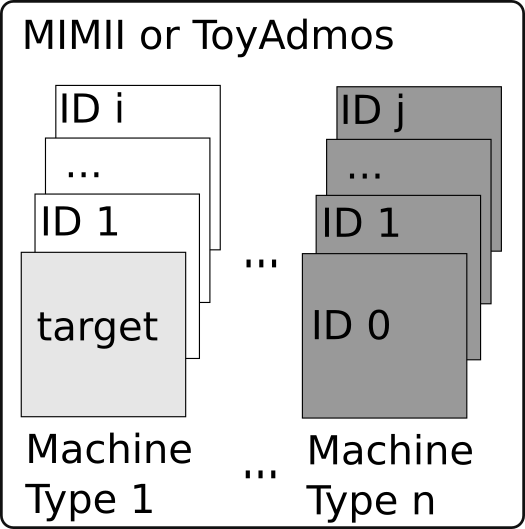}
        \caption{Different Types}
        \label{fig:different}
    \end{subfigure}
    \begin{subfigure}{.15\textwidth}
        \centering
        \includegraphics[width=1\textwidth]{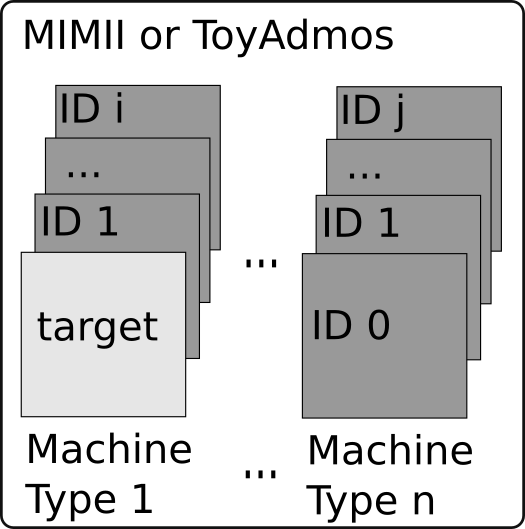}
        \caption{All Types}
        \label{fig:all}
    \end{subfigure}
    \caption{Variants of proxy outlier sets with equal recording conditions. The target set contains the normal sounds (light grey). Proxy outliers sets are selected from the remaining machines' training sets (dark grey).}
    \label{fig:combinations}
\end{figure}

\section{Experimental Setup}
The following section gives a more detailed account of the model architecture, the training procedure, the baseline systems which are based on reconstruction error and density estimation, and the evaluation method.

\subsection{Network Architecture}
We choose the model architecture introduced by Koutini et al.~\cite{Koutini_2019}, a receptive-field-regularized, fully convolutional, residual network (ResNet)~\cite{ResNet}, which has been successfully adopted for various audio-related classification tasks~\cite{Koutini_eusipco, Koutini_emotion}.
The model consists of three stages with four residual blocks each.  The first stage is preceded by a convolutional layer with 64 filters of kernel size $5 \times 5$. We apply global average pooling after the last layer, which allows us to use inputs of varying sizes. In the first stage, a max-pooling layer with kernel size $2 \times 2$ follows after the first, second, and fourth residual block.  Each residual block consists of two convolutional layers, where we use $64$, $128$, and $256$ filters in the first, second, and third stage, respectively. In the first stage, we use filters of size $3 \times 3$, except for the second convolution of the first and fourth block where we use $1 \times 1$ filters. In the other two stages, we use filters of size $1 \times 1$. A batch normalization layer~\cite{IoffeS15} follows each convolutional layer and we use ReLU activations.

\subsection{Preprocessing, Training \& Inference}
 Following the DCASE 2020 Challenge task 2 baseline system~\cite{Koizumi_2019_0}, we re-sample the audio signals to 16000Hz and compute a mono-channel Short Time Fourier Transform using 1024-sample windows and a hop-size of 512 samples. The re-sampled audio waveform is normalized to unit variance. We weight the resulting power spectrogram with a mel-scaled filterbank of 128 filters and apply the logarithm to dampen large outliers. Training of the model is performed on random snippets of $256$ frames length to minimize the binary cross-entropy loss for 100 epochs using the Adam update rule~\cite{adam} with $beta_1 = 0.9$ and $beta_2=0.99$ and a batch size of 64. Batches are stratified to contain 32 positive and 32 negative samples. During one epoch the network sees all normal samples; proxy outliers are drawn randomly. We set the initial learning rate to $10^{-4}$ and decay it each epoch by a factor of $0.99$. The anomaly score for each test example is obtained by collecting all 256-frame windows of the input, computing the logit score for each of them, and then mean aggregating all the logit scores to a single value.

\subsection{Baselines}
The section below describes the baseline methods based on reconstruction error and density estimation. 

\subsubsection{Reconstruction Error Based}
Following the DCASE2020 Challenge baseline~\cite{Koizumi_2019_0}, we train a symmetric autoencoder with eight fully connected layers to minimize the reconstruction error on spectrogram snippets of 5 frames length on the normal data set. All layers have 128 units, except for the bottleneck layer in the middle, which only has eight units. The test samples' reconstruction error, averaged over the whole sample, is used as anomaly score. Compared to the official baseline system, we found this method to perform slightly better if trained per machine instance, and, for this reason, use it as a new baseline. We apply the preprocessing steps and training procedure described above, but change the batch size to $512$ and the learning rate weigh decay to one to match the training procedure of the DCASE2020 Challenge baseline system.

\subsubsection{Density-Estimation Based}

In addition to the autoencoder baseline we train a Masked Autoregressive Flow (MAF)~\cite{PapamakariosMP17_MAFs} model as described in~\cite{Haunschmid2020}. This approach is inspired by~\cite{Schmidt2019_NoveltyDetection} who used different types of normalizing flows (including MAFs) for novelty detection in industrial time series data. Due to their fast data likelihood estimation MAFs are well suited for anomaly detection. The model learns the distribution of the training data which allows using the negative log-likelihood of an unseen sample as its anomaly score. The anomaly score for each sample at test time is computed as the average over all non-overlapping 4 frame snippets. The reported model has 2048 hidden units per layer, 1 hidden layer per invertible block, and consists of 4 such blocks. The model is trained in an unsupervised fashion (only access to normal samples) but conditioned on the machine ID (resulting in a 41-dimensional one-hot encoded vector $y$). To be able to condition on the machine ID one model was trained on all machines available for training at once. We apply preprocessing steps as described above and train the model as described in~\cite{Haunschmid2020}.

\subsection{Evaluation}
As standard in AD, we use  the Area Under the Receiver Operating Characteristics Curve (AUC), which simultaneously takes true positive rate  and false positive rate into account without fixing a threshold to distinguish between normal and anomalous samples. Let $N_+$ and $N_-$ be the set of anomalous and normal samples, respectively, and $\mathcal{A}$ an anomaly detector. The AUC score is then defined as follows:

$$ AUC =\frac{1}{|N_+| \cdot |N_-|} \sum_{x_- \in N_-} \sum_{x_+ \in N_+} \mathds{1}_{\mathcal{A}(x_+) - \mathcal{A}(x_-) > 0} $$
The AUC can be interpreted as the probability that the anomaly detector $\mathcal{A}$ ranks two randomly selected positive and negative samples correctly.

\section{Results \& Discussion}

\begin{figure*}[pt!]

    \includegraphics[width=1\textwidth]{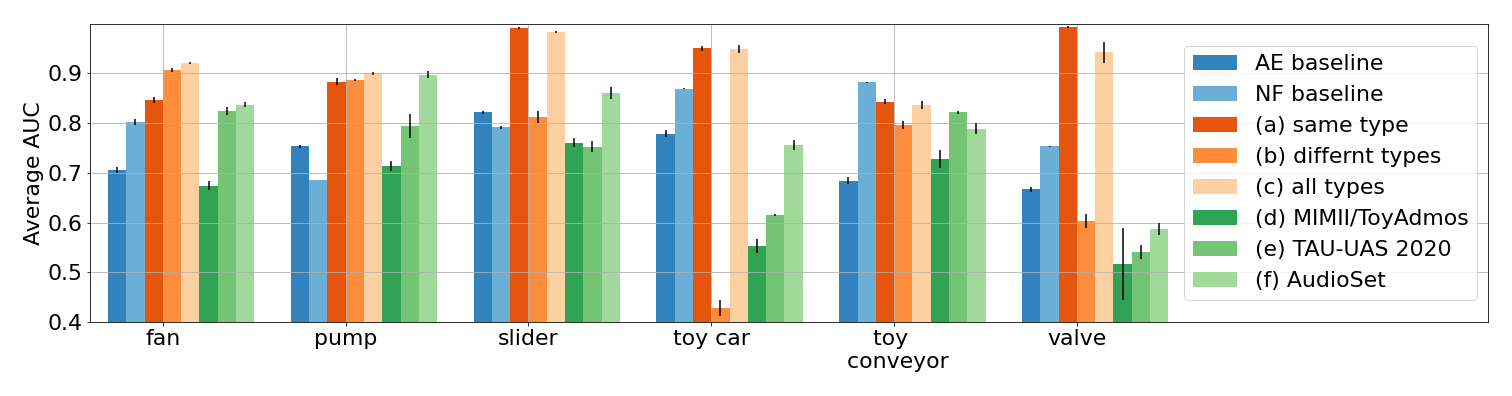}
    \caption{AUC averaged over machine instances per machine type. Blue bars represent the baseline systems. Orange bars are proxy outlier (PO) sets with similar recording conditions, green bars PO sets with mismatching recording conditions. Black error bars represent $\pm 1 $ standard deviation of three runs. }

    \label{fig:results_overview}
\end{figure*}

\begin{figure}[pt!]
    \centering
    \includegraphics[width=0.5\textwidth]{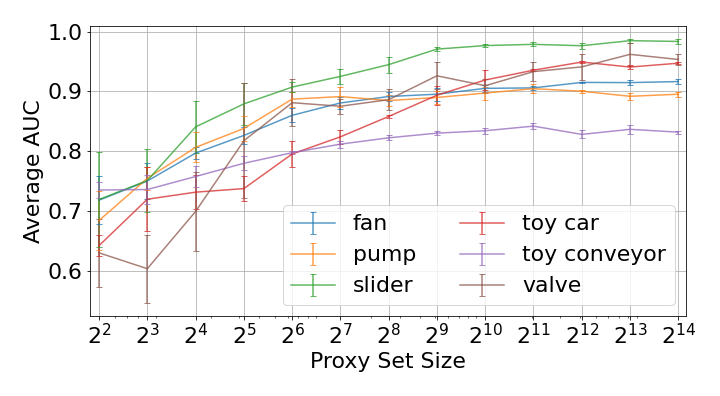}
    \caption{AUC averaged per machine type with increasing number of proxy outlier samples. Proxy outliers are randomly selected from all remaining machine instances with matching recording conditions (c). Error bars represent $\pm 1 $ standard deviation of three runs.}

    \label{fig:sample_size}
\end{figure}

We train three anomaly detectors for each baseline method and each variant of the PO data set. To give an overview of the results, we average the AUC scores per machine type and show the results in Figure~\ref{fig:results_overview}. 

We observe a significant improvement over the baselines (blue bars) if the PO set is composed of samples with matching recording conditions (orange bars), and the set includes machines of the same type as the target machine (dark and light orange). Sounds of the same machine type may be more useful as PO because they are closer to the actual decision boundary.

What stands out in our results is that only using the training set of different machine types leads to a substantial drop in AUC for some machine types, especially for slider, toy car, and valve. This could be explained by the fact that the sounds of different machines are too dissimilar to the target sounds and do not contain diverse content. As a consequence, the classifier might be able to distinguish between normal sounds and POs based on simple statistics and does not learn discriminative features for AD. 

For the PO sets with mismatching recording conditions (green bars), we observe a positive correlation between the improvement over the baseline and the diversity of recording conditions and content. If AudioSet is used as PO set, we notice a significant increase in AUC over the baselines for fan, pump, slider, and toy conveyor and a considerable decrease for toy car and valve (light green). A similar but weaker pattern arises if the PO set is composed of samples from the TAU-UAS dataset (medium green): We observe an improvement for fan, pump, and toy conveyor and no increase or even a deterioration for slider, toy car, and valve. Using MIMII for machine instances in ToyAdmos and vice versa yields a considerable drop in AUC for all machine types except for toy conveyor.

Next, we investigate the role of the PO set size by randomly sampling subsets from all training sets of the remaining machine instances with matching recording conditions. We start with four samples and gradually increase the PO set size to the maximum of 16392 samples (Fig.~\ref{fig:sample_size}). On average, our method outperforms the autoencoder and NF baseline (except for toy conveyor) with only 128 and 512 POs, respectively.

Note that our observations are consistent with the findings by Ruff et al.~\cite{Ruff_2020}, who found that the performance of their outlier detector increased with the diversity of the POs, and only a few POs are necessary to outperform unsupervised baselines. Besides, we identify similarity to the target sounds and matching recording conditions as critical success factors.

\section{Conclusion}

In this study, we show that carefully selected proxies for anomalous sounds can be leveraged to frame the task of anomalous sound detection as a simple classification problem. In our machine condition monitoring experiments, we find that matching recording conditions, similarity to the target sounds, and content diversity make POs exceptionally informative. We also show that relatively few PO samples are needed to outperform unsupervised baselines based on reconstruction error and density estimation.

\section{ACKNOWLEDGMENTS}
We thank Khaled Koutini for making his implementation of the receptive-field-regularized ResNet available.

Parts of this research have been funded within the project AutoDetect (FFG project no. 862019) and by the BMK, BMDW, and the Province of Upper Austria in the frame of the COMET Programme managed by FFG.


\bibliographystyle{IEEEtran}
\bibliography{refs}

\end{sloppy}
\end{document}